# Microstructure Design of Low-Melting-Point Alloy (LMPA)/ Polymer Composites for Dynamic Dry Adhesion Tuning in Soft Gripping

Yaopengxiao Xu[1], Pei-En Chen[2], Wenxiang Xu[1,3,*], Yi Ren[2,*], Wanliang Shan[4,*], and Yang Jiao[1,*]

[1] Materials Science and Engineering, Arizona State University, Tempe AZ 85287

[2] Mechanical Engineering, Arizona State University, Tempe AZ 85287

[3] College of Mechanics and Materials, Hohai University, Nanjing 211100, P.R. China

[4] Mechanical and Aerospace Engineering, Syracuse University, Syracuse, NY 13244

*Corresponding authors: xwxfat@gmail.com (W. X.), yiren@asu.edu (Y. R.), washan@syr.edu (W. S.), yang.jiao.2@asu.edu (Y.J.)

**Abstract**

Tunable dry adhesion is a crucial mechanism in compliant manipulation. The gripping force, mainly originated from the van der Waals' force between the adhesive composite and the object to be gripped, can be controlled by reversibly varying the physical properties (e.g., stiffness) of the composite via external stimuli. The maximal gripping force $F_{max}$ and its tunability depend on, among other factors, the stress distribution on the gripping interface and its fracture dynamics (during detaching), which in turn are determined by the composite microstructure. Here, we present a computational framework for the modeling and design of a class of binary smart composites containing a porous low-melting-point alloy (LMPA) phase and a polymer phase, in order to achieve desirable dynamically tunable dry adhesion. In particular, we employ spatial correlation functions to quantify, model and represent the complex bi-continuous microstructure of the composites, from which a wide spectrum of realistic virtual 3D composite microstructures can be generated using stochastic optimization. A recently developed volume-compensated lattice-particle (VCLP) method is then employed to model the dynamic interfacial fracture process to compute $F_{max}$ for different composite microstructures. We focus on the interface defect tuning (IDT) mechanism for dry adhesion tuning enabled by the composite, in which the thermal expansion of the LMPA phase due to Joule heating initializes small cracks on the adhesion interface, subsequently causing the detachment of the gripper from the object due to interfacial fracture. We find that for an optimal microstructure among the ones studied here, a 10-fold dynamic tuning of $F_{max}$ before and after the thermal expansion of the LMPA phase can be achieved. Our computational results can provide valuable guidance for experimental fabrication of the LMPA-polymer composites.

# 1. Introduction

Soft (compliant) grippers have become a crucial component in many soft robotics systems for a wide range of applications due to their dynamic configurability in posture, shape and adhesion strength to handle a wide range of objects with different sizes, shapes and weights [1, 2]. Dynamic dry adhesion tuning (DDAT) is one of the crucial mechanisms in compliant manipulation [3-5]. The gripping force, mainly originated from the van der Waals' forces between the adhesive units (in particular the adhesive composite) and the object to be gripped at the gripper-object interface, can be controlled by reversibly varying the physical properties (e.g., stiffness) of the composite via external stimuli. For example, tuning the stiffness of the adhesive unit of the gripper (i.e., the adhesive composite) can change the stress distribution at the gripper-object interface, which in turns can lead to the initiation and propagation of cracks, and eventually interfacial fracture (i.e., detachment of the gripper and the object). Therefore, a large tunability of dry adhesion is highly desirable for soft gripping, which will enable the gripper to pick up heavy objects and also release light objects.

Recently, the realization of dynamic dry adhesion tuning (DDAT) as well as other soft robotic mechanisms through the development of multifunctional smart composites has been extensively investigated [4-8]. For example, conductive propylene-based elastomers (cPBE) can reversibly change their mechanical rigidity when powered with electrical current. In particular, the material softens due to Joule heating as the temperature exceeds its melting point, which leads to a reversible change of its elastic modulus over one order of magnitude (e.g., from ~ 1 MPa to ~ 40 MPa) [7], which has been used in activation mechanisms for multiple instances of soft grippers [2, 4]. Shape memory materials, such as shape memory polymers (SMP) and shape memory alloys (SMA), which are able to change stiffness through stimulus-induced changes in molecular structure, are another class of promising candidates for realization of DDAT for soft gripping applications [5]. However, most of these smart materials for DDAT still suffer from slow actuation, low robustness, poor reversibility, and sometimes a combination of these, which can be attributed to the nonsystematic, trial-and-error approaches for the development of such composites [6-8].

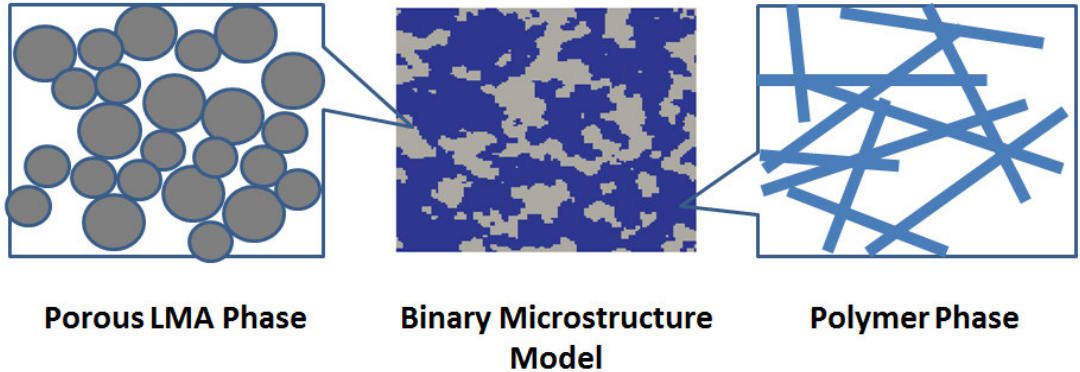

**Figure 1**: Schematic illustration of the porous-LMPA-PDMS composite. The porous LMPA phase is composed of sintered LMPA powders with tunable porosity, which enables tunable physical properties of the phase including thermal expansion, stiffness and density. Both the LMPA phase and the PDMS matrix phase are percolating in 3D, forming an interpenetrating bi-continuous microstructure (middle panel).

Here, we propose a novel hypothetical composite based on the combination of a conductive, rapid phase-change porous low-melting-point alloy (LMPA, e.g., Bi-Pb-Sn-Cd-In-Tl) and a soft, stretchable matrix (e.g., poly (dimethylsiloxane) or PDMS) to resolve those limitations. An illustration of the microstructure of this composite is provided in Fig. 1. Compared to the elastomer-based multifunctional materials, such as CPBEs, the use of LMPA as the conducting phase increases the electrical conductivity of the composite, and allows for dramatic rigidity change when LMPA is melted quickly under Joule heating, while the encapsulating PDMS maintains the pre-molten morphology of the LMPA phase and the integrity of the overall material. In addition, the porous LMPA phase can also reduce the density as well as the overall stiffness of the composite.

More importantly, the porous LMPA phase also enables the interface defect tuning (IDT) mechanism for dynamic dry adhesion tuning due to its large tunable thermal expansion for moderate temperature increase (e.g., ~ 1% in volume, and the thermal expansion can be further tuned by varying the porosity of the LMPA phase). As illustrated in Fig. 2, the large thermal expansion of the LMPA inclusions especially at the gripper-object interface can significantly change the stress distribution on the interface and also lead to initiation of interface cracks, which depends on the degree of Joule heating (and thus, the applied current). The interface cracks can significantly weaken the interface, which in turn can dramatically reduce the maximal gripping force $F_{max}$ associated with the gripper, achieving desired rapid dynamic dry adhesion tuning.

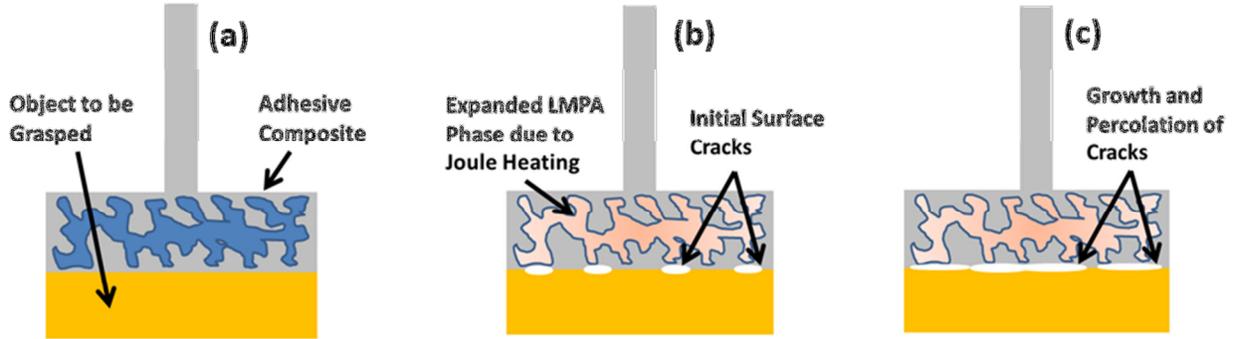

**Figure 2:** Schematic illustration of soft gripping via dynamic dry adhesion tuning (DDAT). (a) The adhesive unit (made of adhesive composite) is attached to the surface of an object to be grasped (shown as the yellow block), forming a perfect interface. The strength of the interface, depending on the van der Waals' forces between the adhesive composite and the object, determines the maximal gripping forces $F_{max}$ of the gripper. (b) The thermal expansion of the LMPA inclusions due to Joule heating induced by electric current leads to initiation of cracks at the interface, which can significantly weaken the interface. (c) The growth and percolation of interface cracks leads to interfacial fracture, which results in a significantly reduced $F_{max}$ and subsequent detachment of the object.

In this paper, we develop a computational framework for the modeling and design of the microstructure of the LMPA-PDMS composites in order to achieve desirable dynamically tunable dry adhesion. In particular, we employ spatial correlation functions to quantify, model and represent the complex interpenetrating bi-continuous microstructure of the composites, and use stochastic material reconstruction methods to generate a wide spectrum of realistic virtual 3D composite microstructures based on the correlation functions. In particular, we employ a damped oscillating two-point correlation function, i.e., $S_2(r) = \exp\left(-\frac{r}{a}\right)\cos(wr)$ for quantitatively representing the LMPA composite microstructure and systematically tuning the microstructure by varying the characteristic correlation lengths (e.g., *a* and *w*).

A recently developed volume-compensated lattice-particle (VCLP) method is then employed to model the dynamic interfacial fracture process to compute the maximal gripping force $F_{max}$ for different composite microstructures, both before and after the thermal expansion of the LMPA phase (by turning on and off the electric current). We focus on the interface defect tuning (IDT) mechanism for dry adhesion tuning enabled by the thermal expansion of the LMPA phase due to Joule heating, which initializes small cracks on the adhesion interface, subsequently causing the detachment of the gripper from the object due to interfacial fracture. We find that for an optimal microstructure among the ones studied here, a 10-fold dynamic tuning of $F_{max}$ before and after the thermal expansion of the LMPA phase can be achieved. Our computational results can provide valuable guidance for experimental design and fabrication of the LMPA-polymer composites.

## 2. Methods

### 2.1 Microstructure modeling via spatial correlation functions

In general, the microstructure of the LMPA-PMDS composite can be quantified by specifying the so called "indicator functions" associated with both phases of the material [9]. Consider a composite with volume $V$, which contains two disjoint phase regions (see Fig. 1 for illustration): phase 1 (i.e., the porous LMPA inclusion phase) with a total volume $V_1$ and volume fraction $\varphi_1$, and phase 2 (i.e., the PDMS matrix phase) with a total volume $V_2$ and volume fraction $\varphi_2$. It is straightforward to see that $V_1 \cup V_2 = V$ and $V_1 \cap V_2 = 0$. In the following discussions, we consider phase 1 (LMPA phase) as our phase of interest, and the indicator function $I^{(1)}(x)$ of phase 1 is given by

$$I^{(1)}(x) = \begin{cases} 1, x \in V_1 \\ 0, x \in V_2 \end{cases} \qquad (1)$$

The indicator function $I^{(2)}(x)$ for phase 2 can be defined in a similar fashion, and it is obvious that

$$I^{(1)}(x) + I^{(2)}(x) = 1 \qquad (2)$$

The *n*-point correlation function (or *n*-point probability function) $S_n^{(1)}$ for phase 1 is then defined as [10-12]:

$$S_n^{(1)}(x_1, x_2, \ldots, x_n) = \langle I^{(1)}(x_1) I^{(1)}(x_2) \ldots I^{(1)}(x_n) \rangle \tag{3}$$

where the angular brackets "$\langle \ldots \rangle$" denote ensemble averaging over independent realizations of the composites. In general, the n-point correlation function $S_n$ provides the probability that a randomly selected n-point configuration with all of the points falling into the phase of interest.

The two-point correlation (probability) function $S_2^{(1)}$ can be directly derived from Eq. (3), i.e.,

$$S_2^{(1)}(x_1, x_2) = \langle I^{(1)}(x_1) I^{(1)}(x_2) \rangle \tag{4}$$

If the material system is statistically homogeneous, i.e., the joint probability distributions describing the microstructure are invariant under a translation (shift) of the origin of reference, $S_2^{(1)}$ is a function of the relative displacements between the two points, i.e.,

$$S_2^{(1)}(x_1, x_2) = S_2^{(1)}(x_1 - x_2) = S_2^{(1)}(r) \tag{5}$$

where $r = x_2 - x_1$. If the material system is further statistically isotropic, i.e., the joint probability distributions for the microstructure are invariant under rigid-body rotation of the spatial coordinates, as for the composite microstructure considered in this work, $S_2^{(1)}$ becomes a radial function, depending only on the scalar separation distances,

$$S_2^{(1)}(x_1, x_2) = S_2^{(1)}(|r|) = S_2^{(1)}(r) \tag{6}$$

In the ensuing discussions, we will drop the superscript denoting the phase index in $S_2^{(1)}$ for simplicity. Without further elaboration, $S_2$ always denotes the two-point correlation function for the porous LMPA inclusion phase. Based on its definition, we can easily obtain the limiting values of $S_2$, i.e.,

$$\lim_{r \to 0} S_2(r) = \varphi_1 \text{ and } \lim_{r \to \infty} S_2(r) = \varphi_1^2 \tag{7}$$

Eq. (7) can also be derived from the probability interpretation of $S_2$, i.e., the probability of two random chosen points separated by distance $r$, both falling into the phase of interest. Interested readers are referred to Ref. [9] for detailed discussions of these correlation functions.

**2.2 Stochastic 3D microstructure reconstruction**

In this section, we introduce the general simulated annealing reconstruction procedure, which is originally developed by Yeong and Torquato [13, 14] and recently has been generalized to model a wide spectrum of complex heterogeneous engineering materials [15-27] and even microstructural evolutions [28, 29]. In the original YT framework, the reconstruction problem is formulated as an "energy" minimization problem, in which the energy $E$ is defined as the sum of the squared difference between a set of target correlation functions $f^*(r)$ (either computed from the target microstructure of interest or mathematically constructed) and the corresponding functions $f(r)$ computed from a simulated microstructure, i.e.,

$$E = \sum_{r=0}^{\infty}[f(r) - f^*(r)]^2 \qquad (8)$$

The energy minimization problem is then solved using the simulated annealing method [30, 31]. Specifically, starting from a random initial trial microstructure that contains a fixed number of voxel for each phase as determined by the target correlation functions, the positions of two randomly selected voxels with different phases are switched, which results in a new trial microstructure. The correlation functions for the old and new trial microstructures are computed to obtain the associated energy $E_{old}$ and $E_{new}$. The probability that the new trial microstructure will be accepted to replace the old trial microstructure is given by,

$$P_{acc}(old \to new) = min\left\{1, \exp(\frac{E_{old}}{T})/\exp(\frac{E_{new}}{T})\right\} \qquad (9)$$

where $T$ is an effective temperature used to control the acceptance rate for energy-increasing trial microstructures. In the beginning of the reconstruction, the parameter $T$ is chosen to be relatively high in order to achieve an acceptance probability of approximately 0.5. This allows sufficient exploration of the microstructure space to reduce the risk of being stuck in an un-favored local energy minimum. Then $T$ is gradually reduced according to a prescribed cooling schedule as the simulation proceeds in order to allow the energy to converge to zero. In the current work, we choose an exponential cooling schedule, i.e., $T(t)/T_0 = \gamma^t$, where $0.95 < \gamma < 0.99$. As $T$ gradually decreases, the acceptance probability for energy-increasing trail microstructure will also decrease. Eventually the energy converges to the energy minimum, which is associated with the microstructure that perfectly realizes the specified target correlation functions. In practice, the global energy minimum is extremely difficult to achieve, and we consider the reconstruction is successfully accomplished if the energy $E$ drops below a prescribed tolerance value (e.g., $10^{-6}$).

During the reconstruction process, after a new trial microstructure is generated, the energy $E$ given by Eq. (8) needs to be computed to determine whether the new microstructure should be accepted or not based on the Metropolis rule (9). In a typical reconstruction simulation, a large number of trial microstructures (e.g., several million) will be generated and thus, it is crucial that $E$ can be computed efficiently for the procedure to be computationally feasible. We have developed highly efficient algorithms to compute the energy by re-using the information stored for the old trial microstructure [16, 17]. The basic idea is based on the fact that the new and old trial microstructures only differ by a single pair of switched pixels. The majority of the two-point statistics are not affected by such pixel swapping. Therefore, there is no need to re-compute the entire functions for the new microstructure from scratch, and only the affected values should be identified and updated in order to compute the energy of the new microstructure. The detailed procedures for updating various correlation functions after a pixel swapping are discussed in Ref. [16, 17], which will not be repeated here.

## 2.3 Volume-compensated lattice-particle (VCLP) method for interface fracture simulations

As discussed in Sec. 1, the maximal gripping force $F_{\text{max}}$ of the gripper is determined by the strength of the gripper-object interface (in particular, the dry adhesion force at the interface). The thermal expansion of the LMPA phase leads to initiation of cracks at the interface, which further grow and percolate to cause the interfacial fracture. This can significantly reduce the interface strength, and thus achieving a large dynamic tunability of $F_{max}$. In order to numerically determine $F_{max}$, the interfacial fracture processes need to be accurately modeled. Here, we employ a recently developed volume compensated lattice-particle (VCLP) method to numerically simulate the microstructure-sensitive fracture behavior of the gripper-object interface.

The VCLP method is essentially a full-field integrated constitutive law – cohesive law model [32]. The cohesive law is directly obtained from basic constitutive relationships with analytical solutions. It can handle both linear/nonlinear elastic and elastic-plastic material behaviors (i.e., essential for most alloys and metal-matrix composites) as well as brittle fracture and fatigue behaviors. The VCLP formulation has closed-form solution for cohesive bond potentials, and no trial-and-error calibration (as typically done in peridynamics simulations) is required. The VCLP method has been successfully applied to investigate the structure-property relations of a wide spectrum of complex heterogeneous material systems [33-36]. In the ensuing discussion, we will briefly present the framework for VCLP method. Interested readers are referred to Ref. [32] for detailed discussion of the implementation of the method for different material systems.

The key idea of the VCLP method is that two potential terms are used to describe the strain energy $U_s$ stored in a unit cell, including both a local pair-wise potential $U_{pair}$ between particles and a multi-body potential $U_V$ among non-local particles. The force field between particles $I$ and $J$ not only depends on their relative displacements (i.e., pair-wise potential), but also includes a contribution from all the neighboring particles surrounding them (i.e., non-local multi-body potential). The total energy of a unit cell of this particle system can be expressed as

$$Us = U_{pair} + U_V, \qquad (10)$$

where $Us$ is the total energy of a unit cell in the packing, $U_{pair}$ is the energy associated with the pair-wise potential, and $U_V$ is the energy corresponding to the non-local multi-body potential, which is defined as the volume-change-induced energy (e.g., due to Poisson's effect) and thus, the subscript $V$ is used.

The mathematical forms of the two potential terms depend on the packing lattices of the particles as well as spatial dimensionality. In the case of simple square lattice packing of particles in two dimensions, the strain energy stored in the unit cell possesses the following form:

$$U_s^{square} = \frac{1}{2}\sum_{J=0}^{3} k_1^s (\delta l_{ij})^2 + \frac{1}{2}\sum_{J=4}^{7} k_2^s (\delta l_{ij})^2 \tag{11}$$

where $k_1^s$ and $k_2^s$ are the spring stiffness parameters for springs connecting a center particle with the nearest and the second nearest neighbors, respectively; and the sums are respectively over the nearest and second nearest neighbors; and $\delta l_{ij}$ is the bond elongation. For the plane strain case, the strain energy of volume change of a unit cell is

$$U_v^{square} = \frac{1}{2}V^s T^s (\varepsilon_v^s)^2 = 2R^2 T^s (\varepsilon_{11}^2 + 2\varepsilon_{11}\varepsilon_{22} + \varepsilon_{22}^2), \tag{12}$$

where $R$ is radius of the particle (i.e., half of the edge length of the square) and $T^s$ is the volumetric spring constant. Comparing the components with the stiffness matrix of 2D isotropic homogeneous materials, the potential coefficients can be determined uniquely, i.e.,

$$k_1^s = \frac{2E}{1+v}, k_2^s = \frac{E}{1+v}, T^s = \frac{E(4v-1)}{2(1+v)(1-2v)} \tag{13}$$

The forces between a particle pair (i.e., of a bond connecting a particle pair) can be easily computed as

$$f_{ij} = \begin{cases} k_1^{square} \delta l_{ij} + (\frac{2\sqrt{2}-3}{2}) T^s (\sum_{J=0}^{7} \delta l_{ij}), & J = 0,1,2,3 \\ k_2^{square} \delta l_{ij} + (\frac{2\sqrt{2}-3}{2}) T^s (\sum_{J=0}^{7} \delta l_{ij}) & J = 4,5,6,7 \end{cases} \tag{14}$$

In the fracture model, the failure behavior of materials (in this case, the brittle fracture of the gripper-object interface) is governed by a spring-based critical stretch criterion. The crack initiation and propagation are the natural outcome of the spring breakage. No additional external criteria are required. The failure parameter is defined as

$$s_j^{critical} = \alpha_j \cdot R, \tag{15}$$

where $s_j^{critical}$ and $\alpha_j$ is the critical stretch and failure parameter for the $j$th spring, respectively. $R$ is the particle radius. The spring connected each pair of particle can be treated as permanently broken whenever the stretch length of the bond is longer than the criterion. The readers are referred to Ref. [32] for more detailed discussion of the method.

## 3. Results

### 3.1 Composite microstructure modeling and reconstruction

As discussed in Sec. 1, the LMPA-PDMS composites typically possess a disordered bi-continuous interpenetrating microstructure with both phases percolating (see Fig. 1 for illustration). Here, we employ the two-point correlation function $S_2$ (see Sec. 2.1 for definition) to quantitatively model the complex microstructure. Without loss of generality, we select the LMPA phase as the phase of interest, which possess a volume fraction φ = 0.3. We employ the damped oscillation function to model the composites, which has been shown to be very efficient for capturing the salient features of disordered bi-continuous microstructure [15, 16], i.e.,

$$S_2(r) = \varphi(1-\varphi)\exp\left(-\frac{r}{a}\right)\cos(wr) + \varphi^2, \tag{16}$$

where $a$ is a characteristic correlation length of the material and $w$ determines the spatial correlations within the inclusion phase. Here we fix $a = 20\ \mu m$ and consider various correlation parameter $w$ (i.e., = 0, 0.05, 0.1 and 0.2; see Fig. 3(a)). It can be seen that as $w$ increases from 0, the oscillations in the correlation function also become stronger, indicating stronger local correlations among the LMPA inclusions.

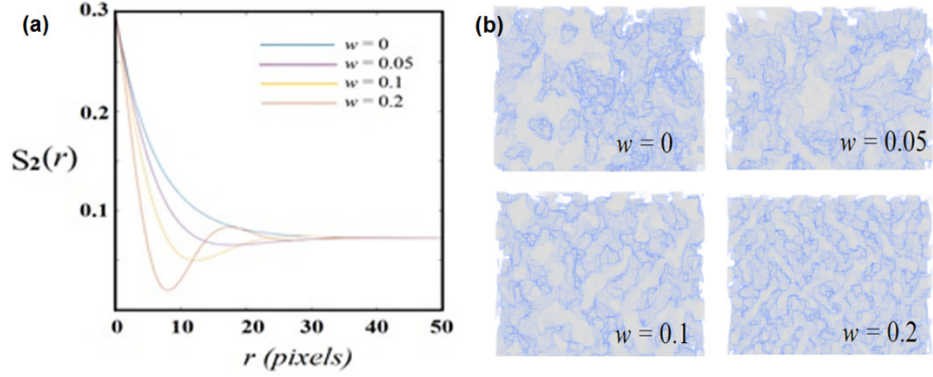

**Figure 3**: (a) Parametrized damped oscillation two-point correlation functions $S_2$ for modeling LMPA-PDMS composites. (b) 3D virtual composite microstructures associated with different correlation functions obtained with stochastic reconstruction. The LMPA phase is shown as light gray (with interface shown in light blue for better visualization effect) and the PDMS matrix is transparent. The linear size of the reconstructed composite microstructure is 100 μm.

Next, we employ the stochastic reconstruction method (see Sec. 2.2 for details) to generate 3D virtual composite microstructures associated with different correlation functions, which are shown in Fig. 3(b). It can be seen that as the correlation parameter $w$ increases from 0, the degree of coarsening of the LMPA phase decreases and the microstructure starts to develop a morphology composed of well-defined ligaments, whose width also decreases as $w$ increases. The ligament morphology contributes to the stronger oscillation in the correlation function for large $w$.

### 3.2 Modeling thermal expansion of LMPA and resulting crack initiation on gripper-object interface

Once the virtual 3D composite microstructures are obtained, we can construct the representative volume element (RVE), which contains the 3D composite material, a homogeneous gripped sample and their interface. Figure 4 show a 3D visualization of the RVE for a composite microstructure associated with $w = 0.2$. Figure 5 shows the side view of the RVEs for all composite microstructures considered in this work as well as the morphology of the composites at the interface.

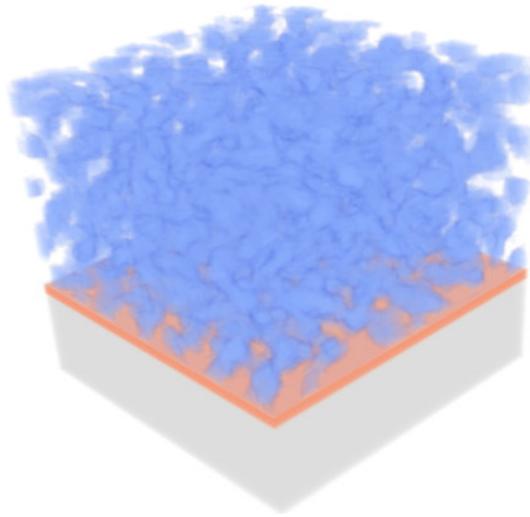

**Figure 4**: 3D visualization of a representative volume element (RVE) for the gripper-composite interface, which includes a cubic composite microstructure associated with $w = 0.2$ and linear dimension of 100 μm, a homogeneous gripped sample (light gray) and the virtual interface (light red). We note that the interface in our simulations is treated as a layer of bonds with distinct mechanical properties from those associated with different materials.

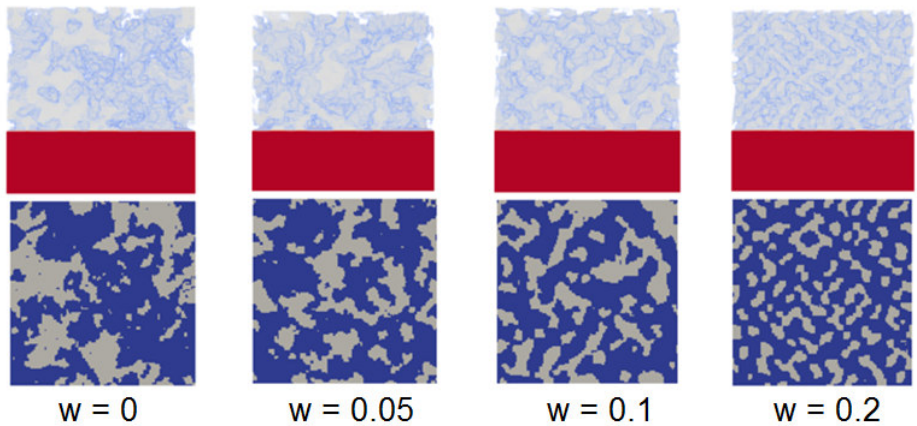

**Figure 5**: Upper panels: side view of the RVEs for all composite microstructures considered in this work. Lower panels: morphology of the composites at the gripper-object interface.

We note that in general the elastic properties of the adhesive composite and the gripped sample can both affect the stress distribution on the interface, and thus, the maximal pulling forces of the gripper. Without loss of generality, we mainly consider the case where the gripped material is thermoplastic with Young's modulus 1 GPa and Poisson's ratio 0.48; but also test other materials with Young's modulus of 100 MPa, and 10 GPa. The Young's modulus of the PDMS and LMPA phases are respectively 1 MPa, 2 GPa; and the Poisson's ratios are 0.3 and 0.45, respectively. The interface between the gripped sample and the top composites is generally weaker than either materials due to the much weaker bonding mechanisms (i.e., van der Waals' force) giving rise to the dry adhesion. As discussed in Sec. 1.3, the key parameter for modeling interfacial fracture is the critical bond elongation, which is chosen as $\alpha = 0.01$ and the Young's modulus for the interface bond is chosen as 0.5 MPa. These parameters are calibrated based on the maximally pulling force of a gripper composed of pure PDMS [5].

Next, an isotropic volumetric thermal strain $\delta$ is applied (0.05% to 0.5%) to the LMPA inclusion phase, which mimics the thermal expansion of the LMPA due to Joule heating, and affects the stress distribution on the gripper-object interface, eventually leading to crack nucleation. We note that we do not explicitly consider the thermos-mechanical constitutive relations of the LMPA phase, which could be also incorporated in the VCLP formulation. Nonetheless, our focus is the to investigate the tunability of the dry adhesion due to thermal expansion of the LMPA phase, which can be very well modeled by applying volumetric thermal strain.

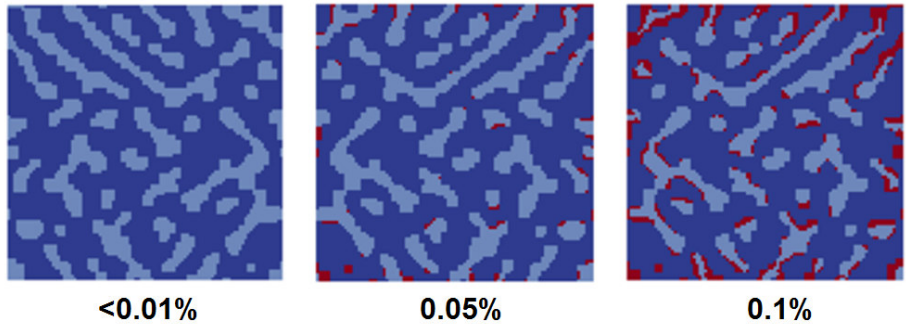

**Figure 6**: Initiation of interface cracks associated with different thermal strains obtained using the VCLP simulations. The linear size of the square interface is 100 μm, which is associated with the composite microstructure with *w* = 0.2. The LMPA and PDMS phases are respectively shown in light and dark blue. The cracks, which are mainly in the vicinity of the LMPA inclusions, are shown in red.

During the thermal expansion process, the thermal strain leads to extrusion of the LMPA phase, initially along the normal direction of the interface. This in turn results in a local tensile state in the vicinity of the LMPA phase at the interface. For sufficiently large thermal expansion (thermal strains), the local tensile stress is large enough to break the (relatively weak) bonds in the vicinity of the LMPA phase, leading to the initiation of interface cracks. As the thermal strain increases, the number and size of the cracks also increase dramatically. Figure 6 shows the initiation of interface cracks associated with different thermal strains, which are obtained using the VCLP simulations. In the subsequent simulations, we will investigate the maximal pulling forces of the gripper for the cases of both perfect interface and cracked interface due to thermal strains.

**3.3 Tuning maximal pulling force via heat-softening of LMPA**

In this section, we investigate the maximal pulling forces $F_{max}$ of the grippers made of different LMPA/PDMS composites, for both a perfect interface and a cracked interface due to thermal expansion of LMPA. The perfect interface condition corresponds to the case that the gripper is attached to the object and forms bond due to dry adhesion. As the gripper attempts to pick up the object, the interface is under uni-axial tensile state. If the object is too heavy, the tensile forces are large enough to break the interface bond, causing the fracture of the interface, i.e., the object cannot be picked up by the gripper. Therefore, the maximal pulling force $F^U_{max}$ for the perfect interface case provides an upper bound on the weight of the objects that can be picked up by the gripper.

On the other hand, once the object is moved to a desired location, it should be released from the gripper. This is achieved by dynamically tuning the dry adhesion forces, i.e., in this case, reducing the $F_{max}$ via significantly weakening the interface bonding by inducing cracks via thermal expansion of the LMPA inclusions. The corresponding $F^L_{max}$ associated with cracked interface provides a lower bound on the weight of the objects that can be released by the gripper. Therefore, a larger ratio of $F^U_{max}/F^L_{max}$ is highly desirable for the gripper to handle objects with a wide range of weights.

In our VCLP simulations, we will numerically obtain both $F^U_{max}$ and $F^L_{max}$ by investigating respectively the dynamic fracture of the perfect interface and the cracked interface due to thermal strains of LMPA phases. In particular, a uniaxial load is applied to the upper and lower boundaries of the RVE of the gripper-object interface, under the quasi-static displacement boundary condition. Periodic boundary conditions are used in the lateral directions. In particular, we apply the load incrementally, with a strain rate $d\epsilon/dt = 10^{-7}$ per loading step and up to $N = 40,000$ loading steps are used. After each loading step, the stress and strain distributions in the RVE are computed and each bond (particularly the interface bonds) is checked for possible breaking. If the critical elongation of the bond is greater than the threshold [c.f. Eq. (15)], this bond (i.e., the connection between two materials points) is removed from the system, which corresponds to the nucleation of a micro-void. This will lead to a significant change of the stress/strain distributions in the next loading step, and typically results in a higher stress state in the vicinity of the micro void, which further leads to additional bond breaking. The reaction force $F$ (and the uniaxial tensile force measured at that upper and lower boundaries of the RVE) is computed at each loading step. The material is considered to fail when an abrupt drop is observed in $F$ (see Fig. 7) The fracture strength can also be computed from the maximum reaction force, i.e., $\sigma_f = F_{max}/A$, where $A$ is the cross-sectional area of the RVE.

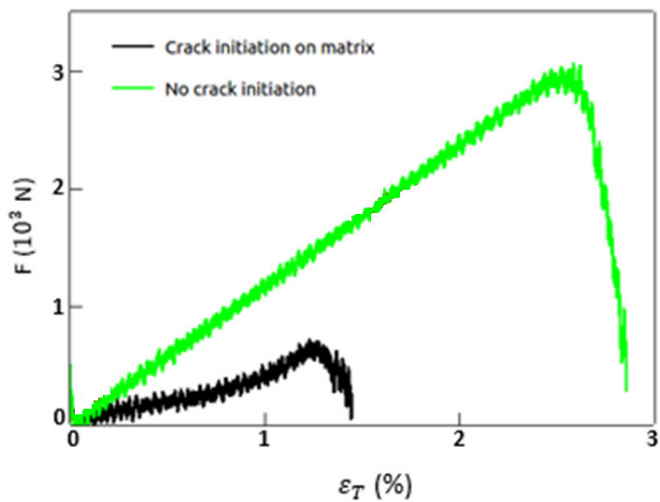

**Figure 7**: Reaction force as a function of tensile strain $\varepsilon_T$ for the prefect interface case (green curve) and the cracked interface case associated with a 0.5% thermal strain (black curve). The gripper is composed of a LMPA/PDMS composite with $w = 0.2$ (see Fig. 8).

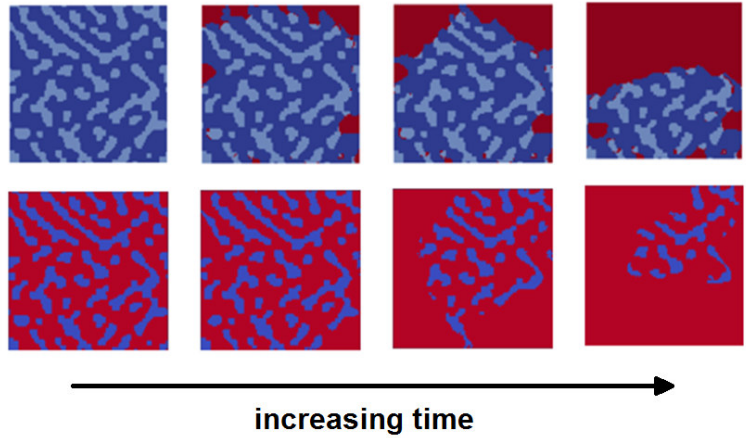

**increasing time**

**Figure 8**: Evolution of the interfacial cracks during dynamic fracture of the gripper-object interface for both the perfect interface (upper panels) and the cracked interface (lower panels) due to a 0.5% volumetric thermal strain of the LMPA inclusions. The microstructure is associated with $w = 0.2$ and has a linear dimension of 100 μm.

Figure 7 shows the reaction forces as a function of the applied tensile strain, for both the perfect interface case (green curve) and the cracked interface case (i.e., associated with a 0.5% thermal strain, black curve) for the gripper composed of a LMPA/PDMS composite with $w = 0.2$. It can be clearly seen that in both cases, the interfacial failure is brittle in nature. The corresponding maximal pulling forces $F^U_{max}$ and $F^L_{max}$ (associated with the perfect and cracked interfaces, respectively) are computed as the largest reaction force for all tensile strains. It can be clearly seen that $F^U_{max}/F^L_{max} \sim 6$, indicating large dynamic tunability for the dry adhesion forces. Figure 8 shows the evolution of the interface cracks during dynamic fracture of the gripper-object interface for both the perfect interface (upper panels) and the cracked interface (lower panels) due to a 0.5% volumetric thermal strain of the LMPA inclusions.

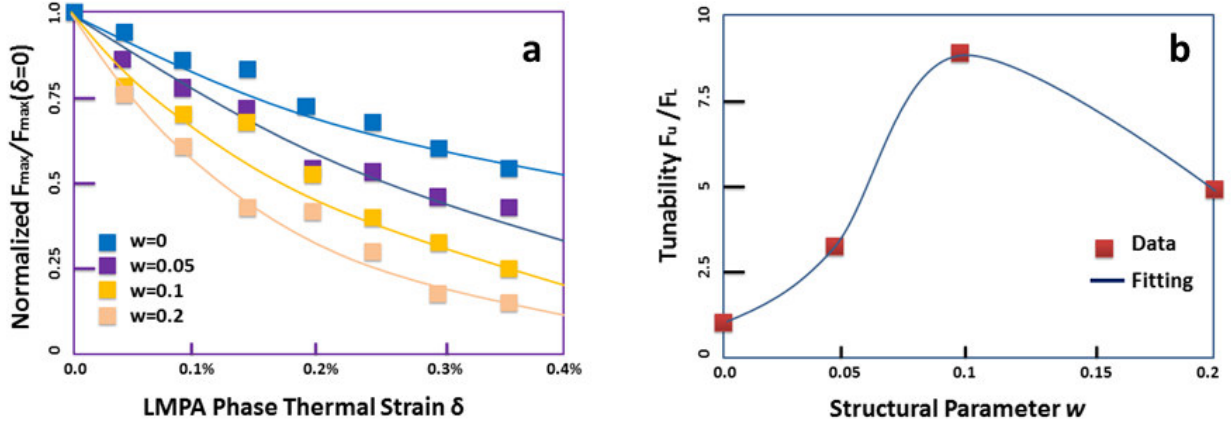

**Figure 9**: (a) $F^L_{max}/F^U_{max}$ as a function of δ up to 0.5% volumetric thermal strain for different composite microstructures. (b) The contrast of $F^U_{max}$ and $F^L_{max}$ at δ = 0.5% for different composite microstructures.

Following the same procedure, we investigate the maximal pulling forces associated with the perfect interface and the cracked interface as a function of thermal strains δ, for different microstructures generated in Sec. 3.1. Figure 9(a) shows the ratio $F^L_{max}/F^U_{max}$ as a function of δ up to 0.5% volumetric thermal strain. We note that $F^U_{max}$ is also the maximal pulling force associated with zero thermal strain, i.e., δ = 0. It can be seen that for a given composite microstructure, $F^L_{max}$ monotonically decreases as the thermal strain increases, and achieves the minimal value at δ = 0.5%, at which the PDMS matrix is almost completely detached from the gripped object due to the extrusion of the LMPA phase at the interface. Therefore, for a given composite microstructure, the maximal tunability (as quantified by the ratio $F^L_{max}/F^U_{max}$) is achieved for δ = 0.5%.

Figure 9(b) graphically shows the contrast of $F^U_{max}$ and $F^L_{max}$ at δ = 0.5% for different composite microstructures. It can be seen that both $F^U_{max}$ and $F^L_{max}$ decrease as the correlation parameter *w* increases from zero. Interestingly, it can be seen that an optimal tunability (i.e., largest contrast between $F^U_{max}$ and $F^L_{max}$) as quantified by $F^L_{max}/F^U_{max}$ is achieved by the microstructure with *w* = 0.1. This is because as *w* increases, the microstructure starts to develop a morphology composed of finer and finer ligaments (see Fig. 5). The finer ligaments lead to stronger stress concentration effects, which facilitate crack initiation and growth at lower tensile loading level. However, if the stress concentration is so strong, as in the case of very fine ligaments associated with $w = 0.2$, $F^U_{max}$ for the perfect interface is significantly reduced so that the tunability $F^L_{max}/F^U_{max}$ is also reduced.

We also compute $F^U_{max}$ and $F^L_{max}$ at $\delta = 0.5\%$ associated with the optimal microstructure (with $w = 0.1$) but for different gripped materials, with Young's modulus of 100 MPa, and 10 GPa. The $F^U_{max}$ for the softer and harder object are respectively 32 kPa and 23 kPa, and $F^L_{max}(\delta = 0.5\%)$ are respectively 5.6 kPa and 3.1 kPa. The increase of the maximal pulling force for the softer material is mainly due to the resulting more homogeneous stress distribution on the interface. However, it can be seen that a large tunability of the dry adhesion is maintain even the material property vary for two orders of magnitude.

## 4. Conclusions and Discussion

In summary, we have developed a computational framework for the modeling and design of the microstructure of the LMPA-PDMS composites in order to optimize dynamically tunable dry adhesion of compliant grippers made of the composites. In this framework, we have employed spatial correlation functions to quantify, model and represent the complex interpenetrating bi-continuous microstructure of the composites, and utilize stochastic material reconstruction methods to generate a wide spectrum of realistic virtual 3D composite microstructures based on the correlation functions. In addition, we have employed a recently developed volume-compensated lattice-particle (VCLP) method to model the dynamic interfacial fracture process to compute the maximal gripping force $F_{max}$ for different composite microstructures, both before and after the thermal expansion of the LMPA phase, respectively resulting in perfect and cracked gripper-object interfaces. We have focused on the interface defect tuning (IDT) mechanism for dry adhesion tuning enabled by the thermal expansion of the LMPA phase due to Joule heating, which initializes small cracks on the adhesion interface, subsequently causing the detachment of the gripper from the object due to interfacial fracture. We find that for an optimal microstructure among the ones studied here, an approximately 10-fold dynamic tuning of $F_{max}$ before and after the thermal expansion of the LMPA phase can be achieved, and the tunability is not sensitive to the type of materials to be gripped. Our computational results can provide valuable guidance for experimental design and fabrication of the LMPA-polymer composites.

We also note that the assumption that the gripper and object forms a perfect interface before thermal expansion of the LMPA phase might not be true in general in actual experiments. This is because the surface of the gripper and the object to be grasped usually are not perfectly smooth and might contain dust particles of various sizes. Therefore, even before the thermal expansion of the LMPA phase, the gripper-object might already contain pre-cracks, which will further reduce the maximal pulling forces. Nonetheless, we expect that the insights obtained in this computational work, e.g., the mechanisms of interfacial fracture, effects of microstructure on dry adhesion tunability etc. are general and should also apply to the actual experimental situations.

**Acknowledgement:** This work is partially supported by the National Science Foundation under Grant No. 1916878. W. Shan acknowledge the support of Syracuse University through startup funds.